\documentclass[aps,graphicx,twocolumn]{revtex4}

\usepackage{graphicx}

\begin{document}

\title{Frequency-tuning induced state transfer in optical microcavities}

\author{Xu-sheng Xu$^{1*}$, Hao Zhang$^{1}$\footnote{These two authors contributed equally to this work.}, Xiang-Yu Kong$^{1}$, Min Wang$^{1}$, and Gui-Lu Long$^{1,2,3}$\footnote{gllong@tsinghua.edu.cn}}

\address{$^1$State Key Laboratory of Low-Dimensional Quantum Physics, Department of Physics, Tsinghua University, Beijing 100084, China\\
$^2$Beijing Information Science and Technology National Research Center, Beijing 100084, China\\
$^3$Beijing Academy of Quantum Information Sciences, Beijing 100193, China}

\date{\today}

\begin{abstract}
Quantum state transfer in optical microcavities plays an important role in quantum information processing, and is essential in many optical devices, such as optical frequency converter and diode. Existing schemes are effective and realized by tuning the coupling strengths between modes. However, such approaches are severely restricted due to the small amount of strength that can be tuned and the difficulty to perform the tuning in some situations, such as on-chip microcavity system. Here, we propose a novel approach that realizes the state transfer between different modes in optical microcavities by tuning the frequency of an intermediate mode. We showed that for typical functions of frequency-tuning, such as linear and periodic functions, the state transfer can be realized successfully with different features. To optimize the process, we use gradient descent technique to find an optimal tuning function for the fast and perfect state transfer. We also showed that our approach has significant nonreciprocity with appropriate tuning variables, where one can unidirectionally transfer a state from one mode to another, but the inverse direction transfer is forbidden. This work provides an effective method for controlling the multimode interactions in on-chip optical microcavities via simple operations and it has practical applications in all-optical devices.
\end{abstract}

\maketitle

\section{Introduction}

As an important fundamental task, state transfer is widely studied in atomic, optical physics and quantum information for its indispensable role in building optical and quantum devices, such as optical transistor \cite{transistor1,transistor2}, frequency conversion \cite{conversion1} and quantum interface \cite{interface1,interface2,JICiracPRL1997,markprx}. In atomic system, the typical approaches for realizing state transfer are the rapid adiabatic passage \cite{NVVitanovARPC2001} for two-level, the stimulated Raman adiabatic passage \cite{KBergmannRMP1998} for excited state assisted three-level $\Lambda$ quantum systems and their optimized shortcuts to adiabaticity technique\cite{DGORMPSTA,MVBerry2009,XChenPRL20102,ABaksicPRL2016,
YHChenpra,YLiangPRA2015,XKSongNJP2016}.

Optical microcavity, which can effectively enhance the interaction between light and matters \cite{Vahala}, is a good platform for studying optical physics and useful applications. For instance, some interesting physics, such as parity-time-symmetry \cite{pt1,pt2,pt3}, chaos \cite{chao1,chao2} and nonreciprocity \cite{optomnonre1,optomnonre2}, have been demonstrated in microcavities. In applications, microcavities show significant functions for sensing \cite{sensing,cavitysensing1,cavitysensing2,cavitysensing3,cavitysensing4} and processing quantum information \cite{CQED,qip1,qip2,qip3,qip4,qip5,qip6,qip7}.  Photon can be confined in microcavity and can also be transferred to another one via evanescent waves coupling or other interactions.  Realizing state transfer between microcavities is important for making microcavity a good physical system for quantum information processing and optical devices. In quantum computing, all optical microcavity coupling lattice structure can be used for performing boson sampling \cite{boson} and microcavity can also be considered as a quantum bus to connect solid qubits for building quantum computer. To make all-optical device, such as transistor \cite{transistor1,transistor2} and router \cite{router}, the target is achieved by performing the state transfer between microcavities successfully. Some effective protocols for state transfer between optical modes are reported with adiabatic methods \cite{LTianPRL2012,YDWangPRL2012,YDWangNJP}, nonadiabatic approaches \cite{YDWangNJP} and shortcuts to adiabaticity  technique. \cite{HZhangOE,XZhouLPL}. By using optomechanical interactions \cite{MAspelmeyerRMP2014,Long1,Long2,Long3,
CHDongscience,Kuzykpra17,optomaqip,optomcooling}, the protocols are completed successfully by tuning coupling strengths very well to satisfy technique constraints.

When we consider the situation of state transfer in on-chip all optical microcavity system, the coupling strength tuning becomes difficult. To solve this problem, in this paper, we proposed an approach to realize the state transfer task between separated modes in optical microcavities via frequency-tuning. In our protocol, we assume that all the coupling strengths are constant and tune the frequency of intermediate microcavity to control the interactions. With linear and periodic tuning, one can transfer the state from the initial cavity mode to the target successfully. To achieve the faster frequency-tuning induced state transfer (FIST) with high fidelity, we use gradient descent to optimize result and acquire an optimal tuning function. Our protocol also shows an significant nonreciprocity in appropriate area of parameters. Good experimental feasibility and interesting features of our work provides potential applications in quantum computing and optical devices.

This article is organized as follows:  We describe the basic model of multimode interactions in optical microcavities in Sec.~\ref{sec2}. In Sec.~\ref{sec3}, we show how to realize the FIST via numerically calculation. We discuss the linear, periodic and optimized tuning in Sec.~\ref{sec31}, B and C, respectively. In Sec.~\ref{sec4}, we investigate the nonreciprocity in our model. Discussion and conclusion are given in Sec.~\ref{sec5}.

\begin{figure}[!ht]
\begin{center}
\includegraphics[width=8.5cm,angle=0]{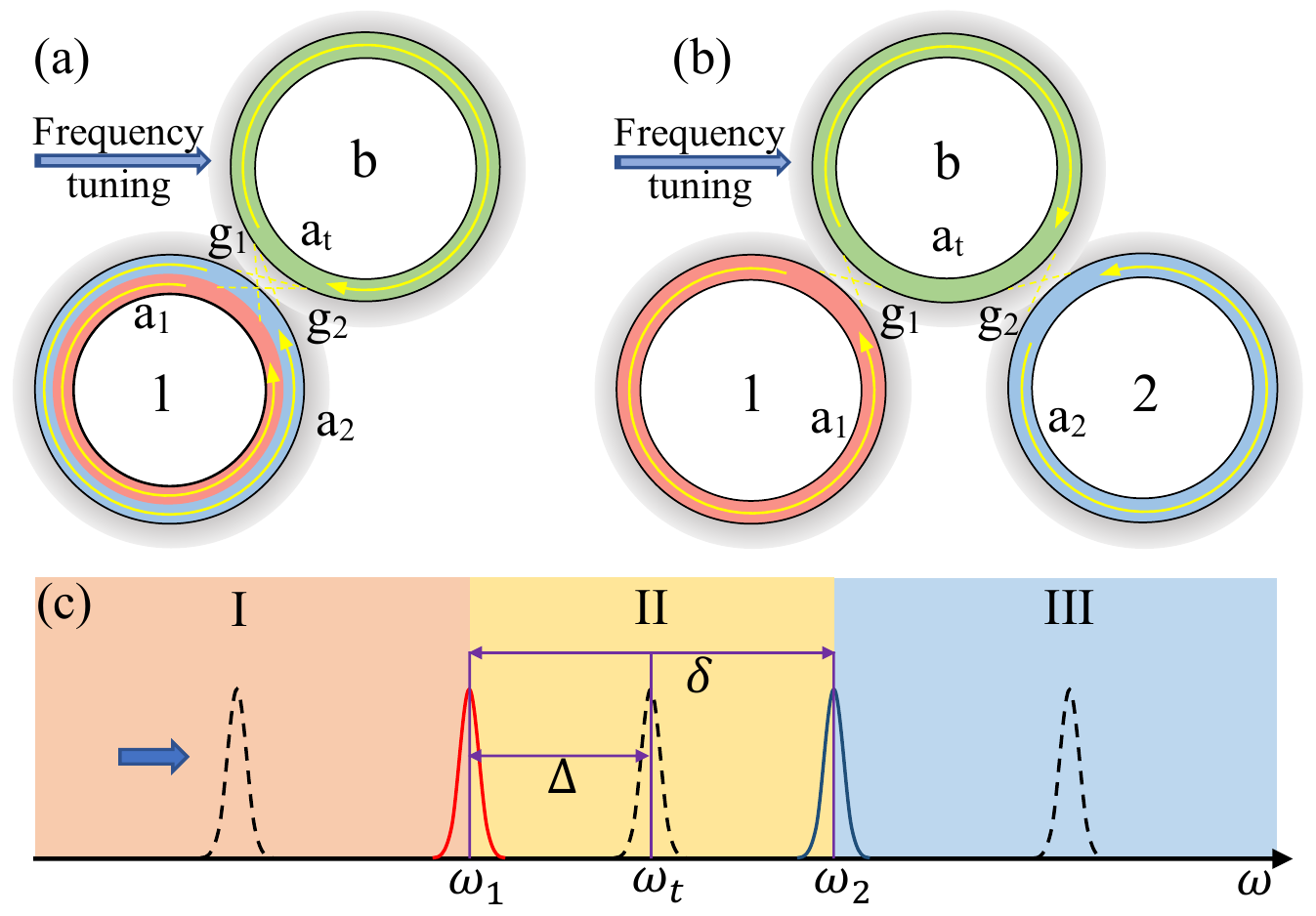}
\caption{Schematic diagram for the model of multimode interactions in optical microcavities. All the modes have very narrow linewidth. A mode in one cavity couples to two different optical modes, (a) in the same cavity, (b) in two different cavities separately.  (c) Resonance frequency tuning of intermediate cavity to induce state transfer. The tuning domain is divided into three parts labelled with I, II and III.}\label{basicmodel} 
\end{center}
\end{figure}

\section{Basic model for multimode interactions in optical microcavities} \label{sec2}
We consider a model of multimode interactions in coupled optical microcavities shown in Fig.~\ref{basicmodel}. The model is universal for situations which two optical modes coupled to the intermediate one and all the modes have very narrow linewidth. Fig.~\ref{basicmodel} (a) gives a setup with nearest neighbour couplings in three optical cavities. We assume that the coupling strengths between corresponding modes are constant and the system can be controlled by tuning the resonance frequency of intermediate mode. Under the rotating frame with unitary transformation $U=\exp[i\omega_{1}t(a_{1}^{\dag}a_{1}+a_{t}^{\dag}a_{t}+a_{2}^{\dag}a_{2})]$, the Hamiltonian is given by
\begin{eqnarray}      \label{eq1}
H_{1}=\delta a_{2}^{\dag} a_{2}+\Delta(t) a_{t}^{\dag}a_{t}+g_{1}a_{1}^{\dag}a_{t}+g_{2}a_{2}^{\dag}a_{t}+\text{H.c.},
\end{eqnarray}
where $a_{i}$ $(a^{\dag}_{i})$ $(i=1,2,t)$ are the annihilation (creation) operators for the corresponding $i$-th mode of the cavity and the corresponding frequency is $\omega_{i}$, respectively. The detunings are $\delta=\omega_{2}-\omega_{1}$ and $\Delta(t)=\omega_{t}-\omega_{1}$. $g_{i}$ $(i=1,2)$ is the coupling strength between modes $a_{t}$ and $a_{i}$. The Hamiltonian can be expressed in Heisenberg equations with $id\vec{a}(t)/dt=M(t)\vec{a}(t)$ and the matrix $M(t)$ is given by
\begin{eqnarray}     \label{eqM}
M(t)=
\left[
\begin{array}{ccc}
0&g_{1}&0\\
g_{1}&\Delta(t)&g_{2}\\
0&g_{2}&\delta\\
\end{array}
\right].
\end{eqnarray}
Here the vector is $\vec{a}(t)=[\hat{a}_{1}(t),\hat{a}_{t}(t),\hat{a}_{2}(t)]^{T}$. To show the results more clearly, we omit the dissipation and noise terms in our calculation due to the very narrow linewidth. In general, the coupling strength between cavities is difficult to be modulated for on-chip sample. Therefore, we keep the coupling strengths constant here and tuning the resonance frequency of intermediate cavity to control the evolution path of system.

\section{FIST between separated modes}\label{sec3}
The frequency-tuning can be realized with different functions. Here we perform the FIST task with two common typical envelopes, i.e. linear and periodic functions, and use the gradient descent technique to optimize the process.

\begin{figure}[ht]
\begin{center}
\includegraphics[width=8.5cm,angle=0]{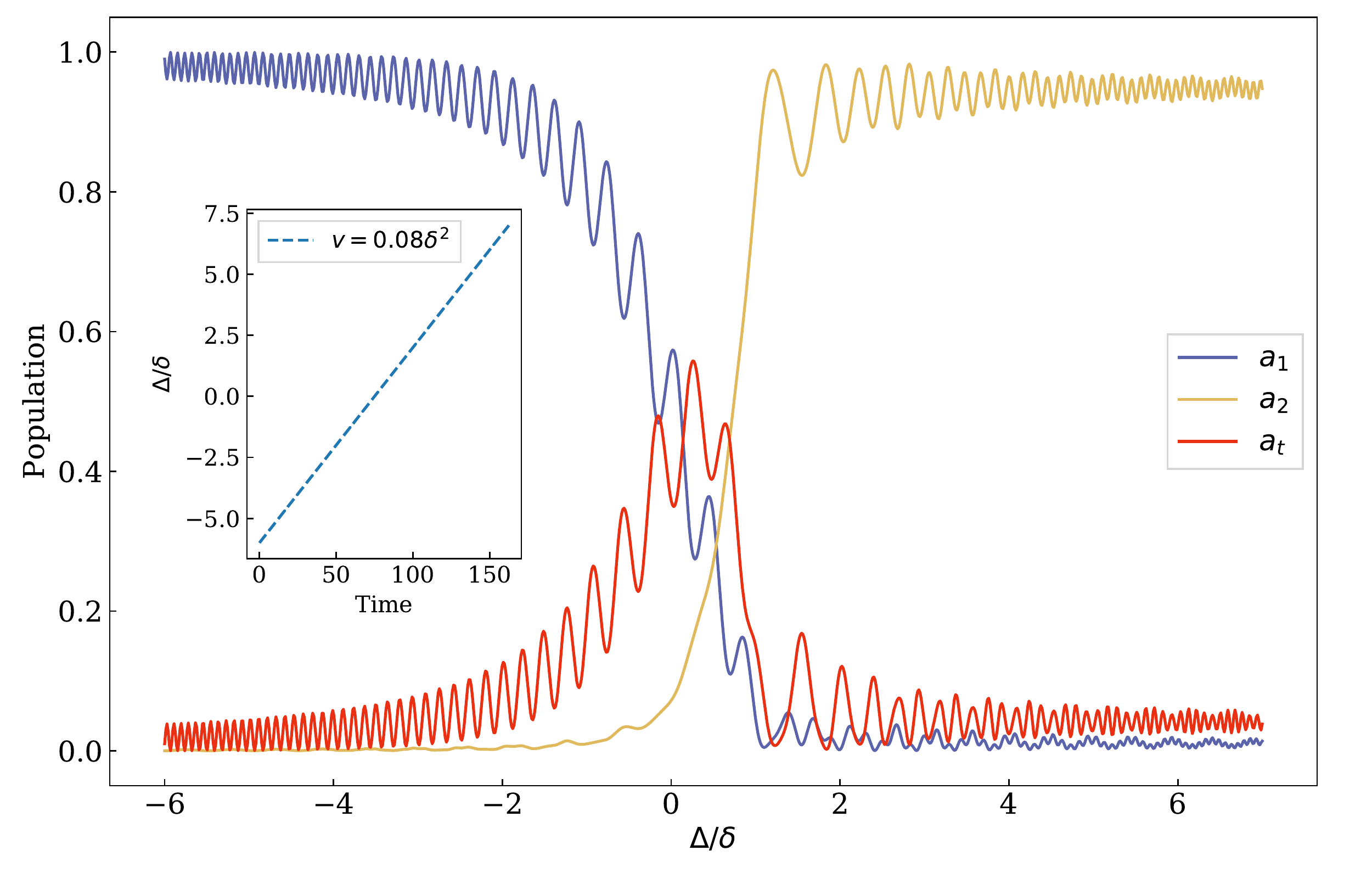}
\caption{The result of FIST between $a_1$ and $a_2$ by linearly tuning the resonance frequency of $a_t$. The speed is chosen with $0.08\delta^2$. The inset is the plot of tuning function and the unit of time $t$ is $\delta^{-1}$.}\label{simulationlinear}
\end{center}
\end{figure}

\subsection{FIST with linear function}\label{sec31}
We assume that the envelope of frequency-tuning is chosen with  $\Delta(t)=vt+\Delta_0$. Here $v$ and $\Delta_0$ are tuning speed and the initial value of detuning, respectively.  At the initial time, the mode $a_{1}$ is stimulated and the other modes are kept in their ground state, i.e. $a_{1}(0)=1$ and $a_{t}(0)=a_{2}(0)=0$. Without loss of generality, we choose the frequency-detuning  of mode $a_{1}$ and $a_{2}$ with $\delta>0$ and sweep the frequency of mode $a_{t}$ from left to right in Fig.~\ref{basicmodel} (c). As the frequency of intermediate mode $a_t$ is swept, the state is transferred from the initial mode to other modes. We numerically simulate the process and show the result in Fig.~\ref{simulationlinear}.  The resonance frequency of intermediate mode is swept from $-6\delta$ to $7\delta$ with a constant speed $0.08\delta^2$, i.e. $\Delta_0=-6\delta$ and $v=0.08\delta^2$. The coupling strength between modes are chosen with $g_{1}=0.6$ and $g_{2}=0.2$ in our simulations. Fig.~\ref{simulationlinear} shows that the population $P$ of mode $a_1$  is transferred to $a_t$ when the frequency of $a_t$ is swept to $a_1$. When the frequency of intermediate mode is moving in domain I of Fig.~\ref{basicmodel} (c), the mode $a_2$ has no effective exchange of population with $a_t$ and almost keeps its initial state due to the large detuning with $a_t$. When the frequency of $a_t$ is swept into domain II (between $a_1$ and $a_2$), $a_t$ exchanges the population with $a_1$ and $a_2$ simultaneously. As the frequency of $a_t$ arrived at domain III, i.e. the right of $a_2$, the mode $a_1$ returns to its ground state and keep the state to the end due to the large detuning with $a_t$. The population of $a_t$ also keep transferring to $a_2$ until the $a_t$ evolves to its ground state. When the detuning $\Delta(t)$ is larger than about $4\delta(t)$, the system evolves to our target state, i.e. $a_{2}(T)\approx1$ and $a_{1}(T)\approx a_{t}(T)\approx 0$. In whole process, one only need to keep a stable tuning speed to get the perfect state transfer after the $a_t$ becomes larger than $4\delta(t)$. The maximal population of $a_2$ is about $P=0.9536$.

\begin{figure}[!ht]
\begin{center}
\includegraphics[width=8.5cm,angle=0]{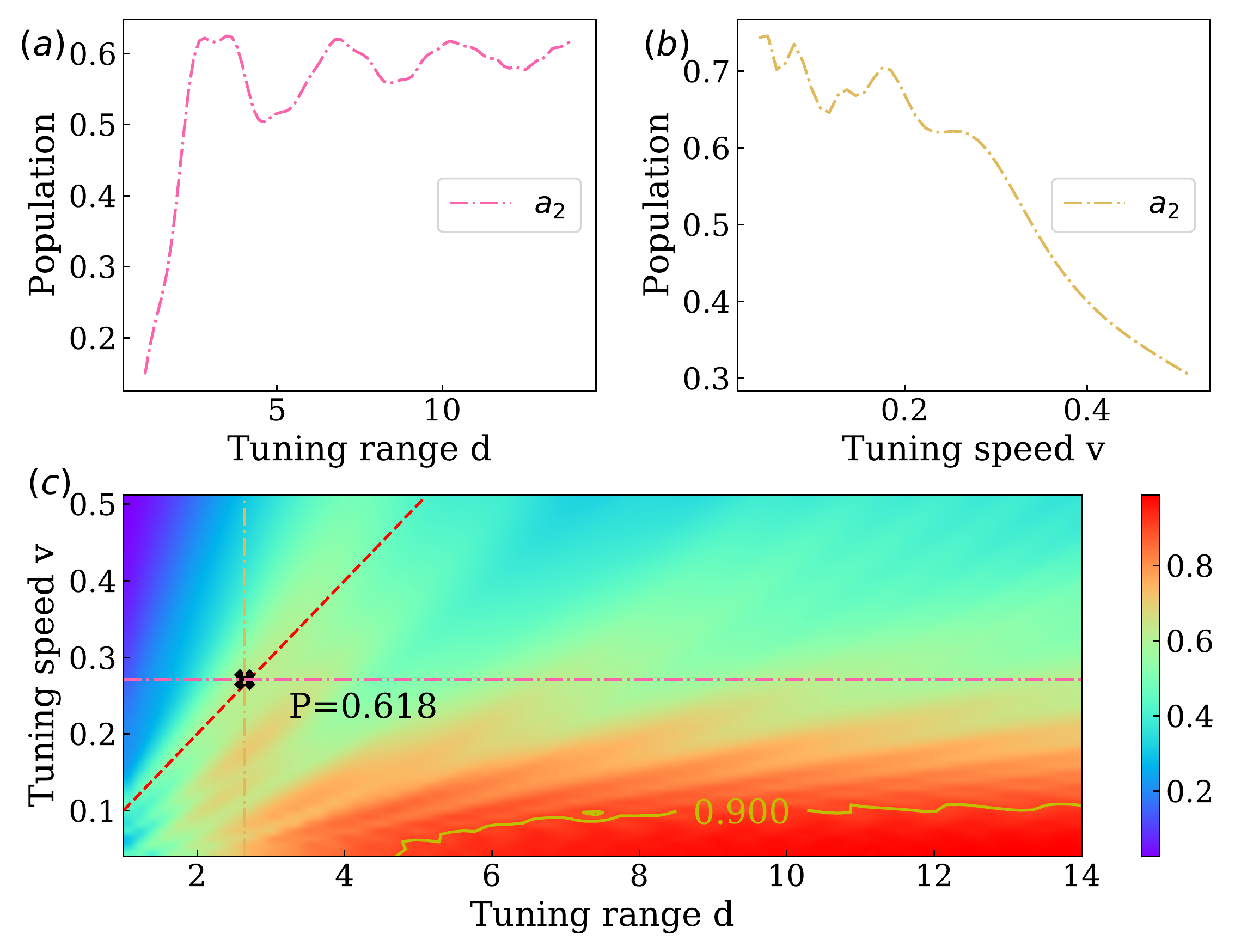}
\caption{Simulation of final population of mode $a_2$ effected by tuning variables. The units of $d$ and $v$ are chosen with $\delta$ and $\delta^2$, respectively. (a) Population $P$ vs tuning range $d$ with $v=0.27\delta^2$. All the $\Delta_0$ are chosen with $\Delta_0=(\delta-d)/2$. (b) Population $P$ vs tuning speed $v$ with $d=2.65\delta$ and $\Delta_0=-0.825\delta$. (c)Population $P$ vs tuning range $d$ and tuning speed $v$. The dashed line is all the points of evolution time with $10\delta^{-1}$.  }\label{popuvd}
\end{center}
\end{figure}

To investigate how FIST is effected by different tuning range $d$ and tuning speed $v$, we show the final population of mode $a_2$ with respective to  $d$ and $v$ in Fig.~\ref{popuvd}. The Fig.~\ref{popuvd} (a) and (b) are the 2-dimensional cross-section analysis of two dot-dashed line in Fig.~\ref{popuvd} (c).  In Fig.~\ref{popuvd} (a), the tuning speed is fixed with $v=0.27\delta^2$. The population is proportional to the tuning range $d$ within a short range less than about $3.0\delta$ and oscillates to a stable value as the range becomes larger. The reason is that system has more time to transfer the state from $a_1$ to $a_2$ as tuning range becomes larger in the domain about $(0,3.0\delta)$. When the tuning range becomes very long, the population has little change. Because the large detuning makes little contribution to state transfer. As shown in Fig.~\ref{popuvd} (b), on the contrary, when the tuning range is kept with $2.65\delta$, the population is inversely proportional to the tuning speed along with a little oscillation, since faster speed makes shorter interaction time.  Fig.~\ref{popuvd} (c) indicates the  clearer conclusion that large tuning range and slow tuning speed are the best tuning manner.

\subsection{FIST with periodic function}\label{sec32}
\begin{figure}[!ht]
\begin{center}
\includegraphics[width=8.5cm,angle=0]{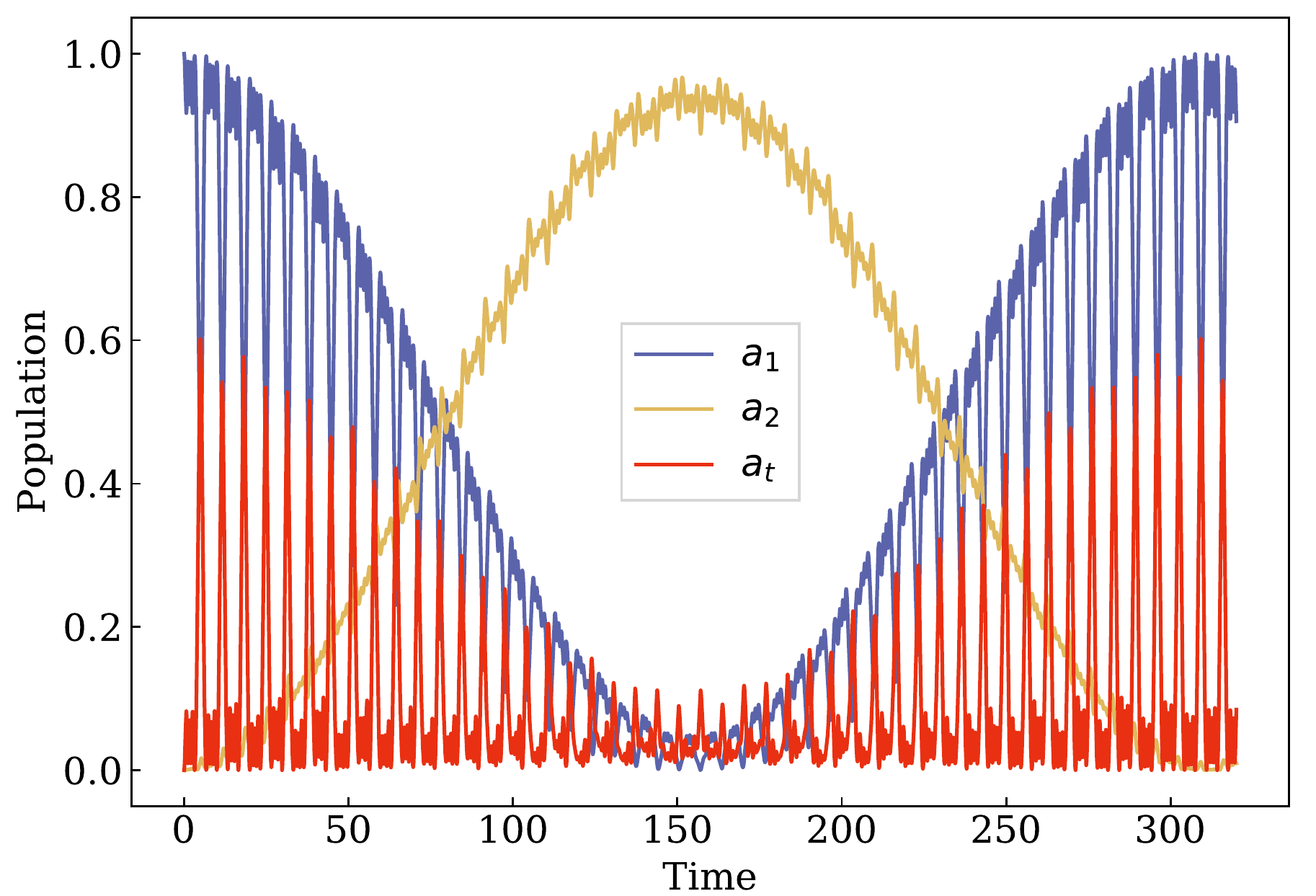}
\caption{The populations change with respective to evolution time via sine tuning function. Lines labelled with $a_1$, $a_2$ and $a_3$ are the populations of corresponding modes, respectively.}\label{popusin}
\end{center}
\end{figure}

Beside the linear tuning, we consider the situation with periodic function. For simplicity, we choose sine function here. The frequency-detuning is described as $\Delta(t)=A[\sin(\Omega t)+c_0]/2$. With  the parameters chosen with $A=9.6$, $\Omega=0.95$ and $c_0=0.5$, the populations of each modes are performed in  Fig.~\ref{popusin}. The envelope curve of populations of $a_1$ and $a_2$ are evolved as a sine function.  The modes $a_1$ and $a_t$ exchange their populations with Rabi frequency $\Omega$, which is the frequency of tuning function. The maximal value of population $P$ of mode $a_2$ in this periodic tuning is $0.9365$. Compared with linear tuning, the population transferred here is realized successfully by controlling the evolution time of tuning function. For instance, in Fig.~\ref{popusin}, the evolution should be ended in around $160\delta^{-1}$. In principle, if the end time is missed, one can wait for the neat periodic time, but the long time will decrease the fidelity of protocol by bring more decoherence.

\subsection{Optimizing the FIST via gradient descent}\label{sec33}
To achieve a perfect and fast state transfer from $a_1$ to $a_3$, we use the gradient descent technique to optimize the FIST and pick an optimal function for frequency tuning. The gradient descent technique is usually used for finding a optimal evolution path for system along the opposite direction of gradient descent. In our model, the evolution equations are given by
\begin{eqnarray}      \label{eqGD1}
a(t)=U(t,0)a(0),
\end{eqnarray}
where the evolution operator is a Dyson series with expression given by $U(t,0)=\exp[\intop^{t}_{0}M(t')dt']$. The time domain is chosen with $t=[0,t_1]$ and the frequency of intermediate mode, considered as parameter to be optimized, is divided into  $n$ discrete constant variables, i.e. $\Delta_I=[\Delta_1,\cdots,\Delta_n]$. Therefore, the operator can be rewritten with
\begin{eqnarray}      \label{eqGD2}
a(t_1)=U(\Delta_1,\cdots,\Delta_n;t_1,0)a(0).
\end{eqnarray}
The target function is chosen with $\mathcal{L}(a,t)=|a_2(t)|^2$ and can be described as 
\begin{eqnarray}      \label{eqGD3}
\mathcal{L}(a,t_1)=\mathcal{L}(\Delta_1,\cdots,\Delta_n;t_1,0).
\end{eqnarray}
Our goal is to optimize above function and get the maximal value. So the gradient of the target function is calculated as $\nabla\mathcal{L}=\frac{\partial\mathcal{L}}{\partial\omega}$. In numerical calculation, the gradient is written approximatively by
\begin{eqnarray}      \label{eqGD4}
\nabla\mathcal{L}=(\nabla\mathcal{L}_1,\cdots,\nabla\mathcal{L}_n),
\end{eqnarray}
where $\nabla\mathcal{L}_i$ is
\begin{eqnarray}      \label{eqGD5}
\nabla\mathcal{L}_i\!=\!\frac{\mathcal{L}(\cdots,\Delta_i+\delta\Delta,\cdots;t_1,0)\!-\!\mathcal{L}(\Delta_1,\cdots,\Delta_n;t_1,0)}{\delta\Delta}.
\end{eqnarray}

\begin{figure}[!ht]
\begin{center}
\includegraphics[width=8.5cm,angle=0]{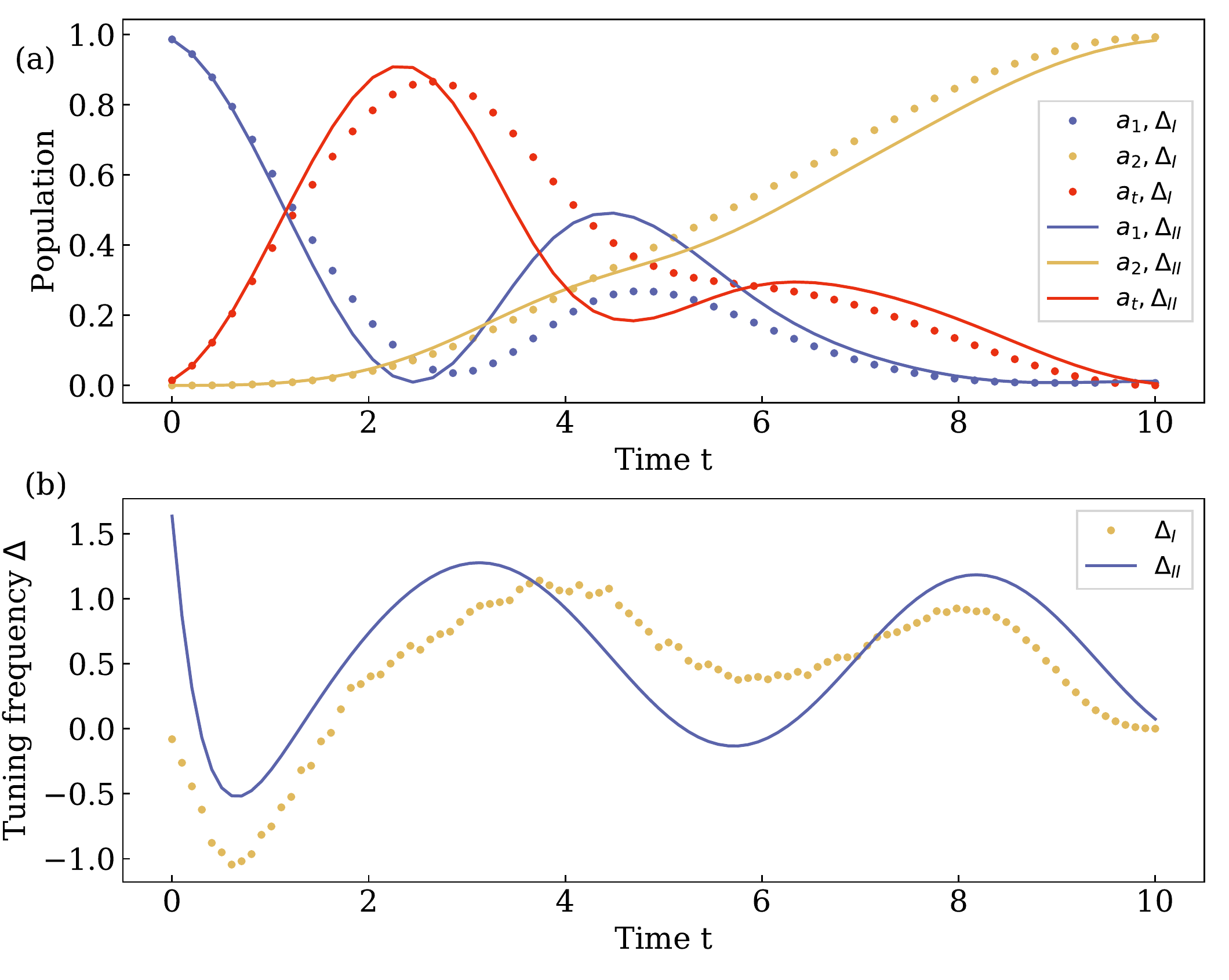}
\caption{Simulation of fast FIST from $a_1$ to $a_2$ by using gradient descent technique. The parameters are the cross point of Fig.~\ref{popuvd} (c) with ($d=2.65\delta$,$v=0.27\delta^2$). (a) The result of optimized  population transfer process. (b) The corresponding optimal tuning function of intermediate mode. The unit of time here is $\delta^{-1}$.}\label{optimize}
\end{center}
\end{figure}

The simulation of optimized fast FIST is shown in Fig.~\ref{optimize}. We plot the transfer results and the tuning functions in Fig.~\ref{optimize} (a) and (b), respectively. In our simulations, we first choose the $\Delta_I(t=0)=0$ and cut the tuning function $\Delta_I(t)$ into discrete $100$ segments which are mutually independent to be optimized via gradient descent algorithm. The results and optimal tuning envelope are plotted respectively in Fig.~\ref{optimize} (a) and (b) with dotted curves. During the evolution time, the population of mode $a_2$ monotonically increases to its maximal value $0.9923$. However, the curves of modes $a_1$ and $a_t$ decrease to its ground state with an oscillating process.

To make the scheme conveniently controlled, according to the envelope of dotted line in Fig.~\ref{optimize} (b), we reasonably give the tuning function with expression as
\begin{eqnarray}      \label{optimalfunc}
\Delta_{II}(t)=A\sin[(\Omega t+\theta)+C_1]*[e^{-\gamma t}+C_2],
\end{eqnarray}
where the parameters $A$, $\Omega$, $\theta$, $\gamma$, $C_1$ and $C_2$ are undetermined coefficients to be optimized. By calculating the state transfer task with the Eq. (\ref{optimalfunc}), the population of mode $a_2$ can be optimized to  $0.9826$ and the parameters are given with $A=-5.555$, $\Omega=1.276$, $\theta=0.564$, $\gamma=1.467$, $C_1=-0.797$ and $C_2=0.119$. The evolutions of populations and tuning functions are plotted in Fig.~\ref{optimize}. The  envelope of Eq. (\ref{optimalfunc}) and the evolution of populations are similar with the dotted curves in each figures, respectively.

In linear tuning shown in Fig.~\ref{popuvd} (c), the maximal population $P$ of mode $a_2$ with evolution time less then $10\delta^{-1}$ is $P=0.618$ labelled with a cross point. With the evolution time $10\delta^{-1}$, the optimized protocol can achieve the population with $P>0.98$. Compared with linear tuning, the state transfer under this protocol can be optimized with a faster evolution path and higher fidelity.

\begin{figure}[!ht]
\begin{center}
\includegraphics[width=8.5cm,angle=0]{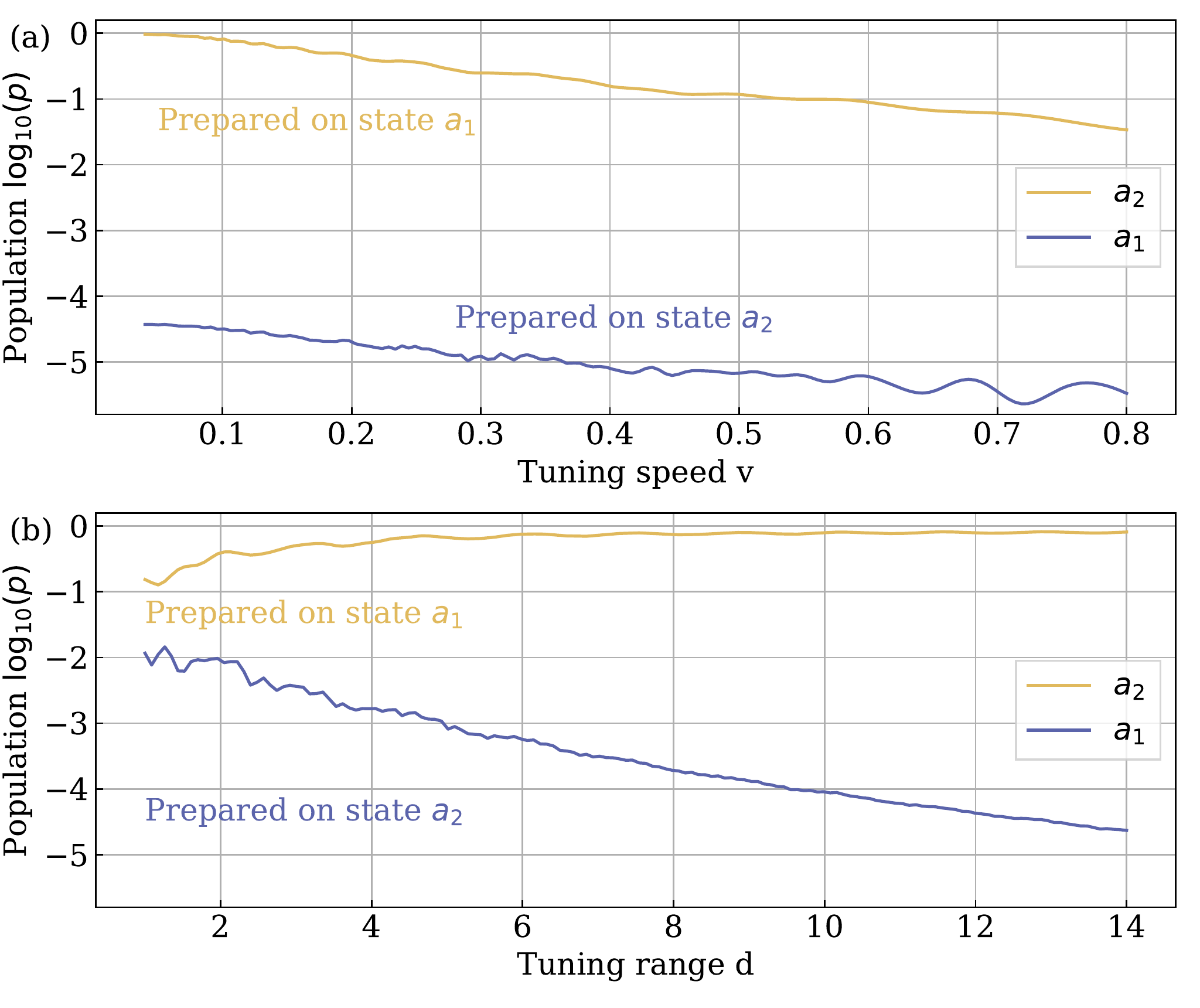}
\caption{Nonreciprocal state transfer between modes $a_1$ and $a_2$. (a) The populations of modes $a_1$  and $a_2$ vs tuning speed. The tuning range is $d=14\delta$. (b)The populations of modes $a_1$  and $a_2$ vs tuning range. The tuning speed is $v=0.1\delta^2$.}\label{Nonreciprocity}
\end{center}
\end{figure}

\section{Nonreciprocity in multimode interactions}\label{sec4}
Our model shows the significant nonreciprocity in state transfer between $a_1$ and $a_2$. For example, when the frequency of intermediate mode is swept from $a_1$ to $a_2$, the state can be transferred from $a_1$ to $a_2$, but it completely failed for $a_2$ to $a_1$. The results of the nonreciprocal state transfer are plotted in Fig.~\ref{Nonreciprocity}  with linear tuning from $a_1$ to $a_2$, i.e. from left to right in Fig.~\ref{basicmodel} (c). We fix the tuning range with $d=14\delta$ and change the tuning speed in Fig.~\ref{Nonreciprocity} (a). The top (bottom) green (blue) line is the final population of $a_2$ ($a_1$) with the initial state is prepared in $a_1$ ($a_2$).  As the speed becomes slower, the  nonreciprocity is more clearly. At very slow speed, the population of $a_2$ transferred from $a_1$ is almost $1$ but the population of inverse transfer process is less than $10^{-4}$. In Fig.~\ref{Nonreciprocity} (b), we investigate the nonreciprocity with respective to tuning range with constant tuning speed $v=0.1\delta^2$. The small populations of both $a_1$ and $a_2$ in different directions in small tuning range indicate that the nonreciprocity is not clearly. When we increase the tuning range $d$, the green and blue line converge towards unit and $10^{-5}$, respectively.

\begin{figure}[!ht]
\begin{center}
\includegraphics[width=8.5cm,angle=0]{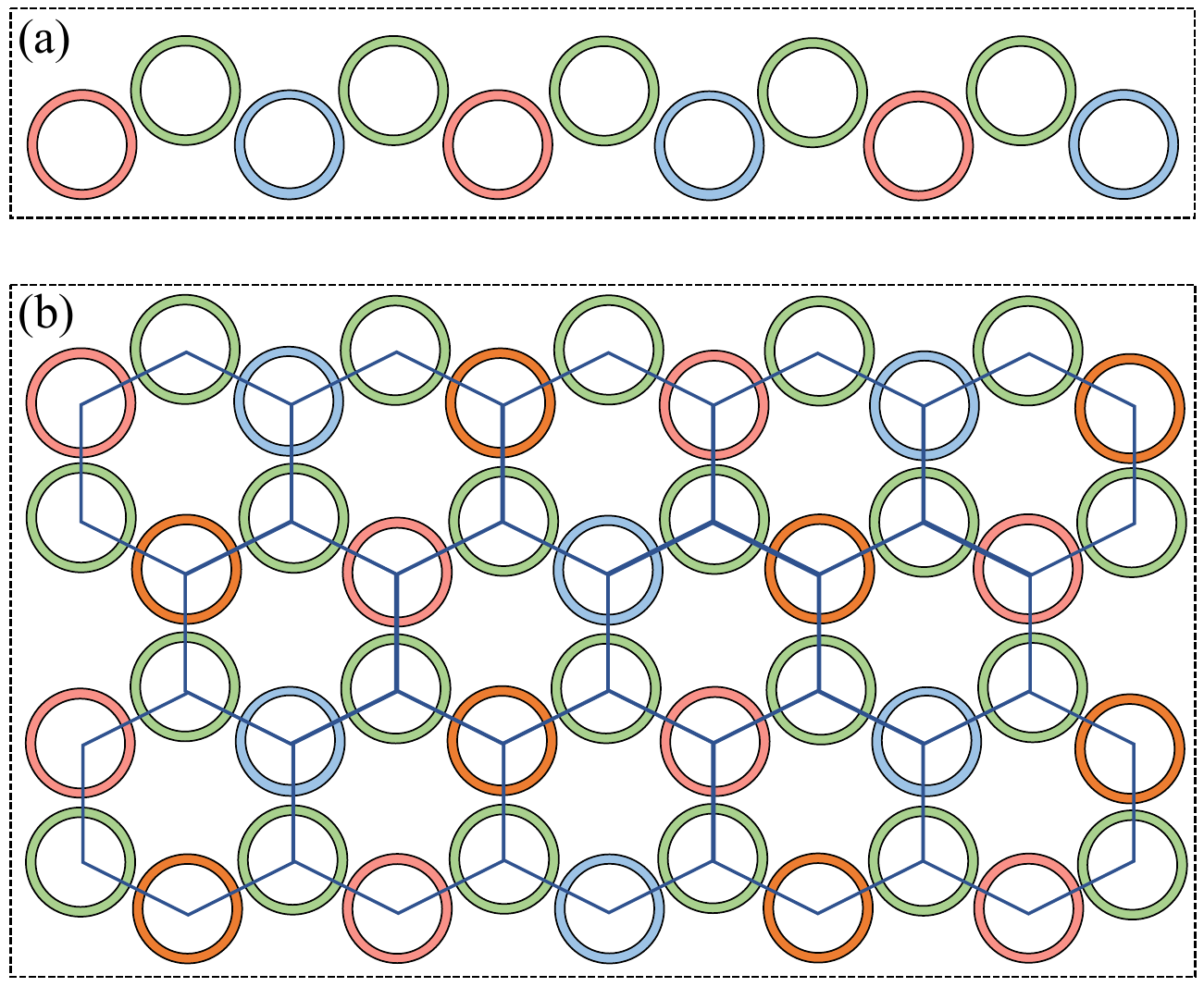}
\caption{All-optical on-chip microcavities structures. (a) One-dimensional microcavity array. (b)Two-dimensional optical microcavity lattice.}\label{network}
\end{center}
\end{figure}

\section{Discussion and Conclusion}\label{sec5}
All the results shown above are considered in all optical cavities system. Actually, our model is a universal  approach for multimodes interaction system, such as all mechanical phonon modes or the photon-phonon interactions. For example, the direct interaction between phonons is difficult. So one can transfer the state from one mechanical resonator to another one via an intermediate optical cavity mode \cite{Kuzykpra17,HZhangOE}. Beside the coupling strength tuning way, one can use frequency tuning approach described here to control the interactions.

To perform more interesting applications, our model can be extended to design one-dimensional microcavity array and two-dimensional microcavity lattice. The detailed schematic diagrams are shown in Fig.~\ref{network}.  An array is built by coupling $n+1$ microcavities  with $n$ tuning cavities in Fig.~\ref{network}(a). The structure can be used for realizing all-optical transistor with more abundant physical tuning. Fig.~\ref{network} (b) shows the two-dimensional lattice structure which designed by connecting one tuning cavity with three storage cavities. Every unit of this structure is a simple optical router. The structure has function for building all-optical on-chip quantum network \cite{JICiracPRL1997,markprx}.

In our model, we always keep the coupling strengths constant with the assumption that the distances between cavities are fixed. The typical corresponding physical system is the on-chip optical microcavity sample. Because the distances between each cavity are difficult to change after complete the fabrication. Our frequency tuning manner is possible.  This is because the frequency of microcavity is sensitive to its shape which can be modulated by some operations, such as temperature \cite{temp1,temp2,temp3}, external forces \cite{mech1,mech2,mech3,mech4,mech5,mech6,mech7} and etc. The above frequency tuning approaches have been realized with high resolution in experiments, but the tuning speed is limited. So the fast tuning way is needed for improving the feasibility of the practical applications.

In conclusion, we have proposed an approach to realize the state transfer between two separated modes in optical microcavities. Our proposal are valid for both two and three microcavities. The FIST can be realized with high fidelity via different tuning manner, i.e. linear and periodic function, of resonance frequency of intermediate mode. To optimize the tuning function, a fast and perfect evolution process is performed by using gradient descent technique. Our proposal also shows the significant nonreciprocity. The state can be transferred successfully in the same direction with frequency tuning and it is fail in the opposite direction. Our work provides an effective approach for controlling the optical mode in on-chip microcavities and has important applications in all-optical devices.

\section*{ACKNOWLEDGMENT}
The authors thank Guo-Qing Qin for helpful discussions.
This work was supported by the National Natural Science Foundation of China (20171311628); National Key Research and Development Program of China (2017YFA0303700); Beijing Advanced Innovation Center for Future Chip (ICFC). H.Z. acknowledges the China Postdoctoral Science Foundation under Grant No.2019M650620.

\end{document}